\documentclass[a4paper,12pt]{article}
\pdfoutput=1 

\usepackage{jheppub} 
\usepackage{bm,latexsym,amsmath,amssymb,amsfonts,mathtools,mathrsfs}
\usepackage{comment}
\usepackage{float}
\usepackage[T1]{fontenc} 

\title{Lifshitz scaling, ringing black holes, and superradiance}

\author[a]{Naritaka Oshita}
\author[a,b]{Niayesh Afshordi}
\author[c,d]{Shinji Mukohyama}

\affiliation[a]{Perimeter Institute, 31 Caroline St, Waterloo, Ontario N2L 2Y5, Canada}
\affiliation[b]{Waterloo Centre for Astrophysics, University of Waterloo, Waterloo, ON, N2L 3G1, Canada}
\affiliation[c]{Center for Gravitational Physics, Yukawa Institute for Theoretical Physics, Kyoto University, 606-8502, Kyoto, Japan}
\affiliation[d]{Kavli Institute for the Physics and Mathematics of the Universe (WPI),
The University of Tokyo, Kashiwa, Chiba 277-8583, Japan}

\emailAdd{naritaka.oshita@gmail.com}
\emailAdd{nafshordi@pitp.ca}
\emailAdd{shinji.mukohyama@yukawa.kyoto-u.ac.jp}

\abstract{We investigate the ringdown waveform and reflectivity of a Lifshitz scalar field around a fixed Schwarzschild black hole. The radial wave equation is modified due to the Lorentz breaking terms, which leads to a diversity of ringdown waveforms. Also, it turns out that Lifshitz waves scattered by the Schwarzschild black hole exhibits superradiance. The Lorentz breaking terms lead to superluminal propagation and high-frequency modes can enter and leave the interior of the Killing horizon where negativity of energy is not prohibited. This allows the Lifshitz waves to carry out additional positive energy to infinity while leaving negative energy inside the Killing horizon, similar to the Penrose process in the ergosphere of a Kerr spacetime. Another interesting phenomenon is emergence of long-lived quasinormal modes, associated with roton-type dispersion relations.  
These effects drastically modify the greybody factor of a microscopic black hole, whose Hawking temperature is comparable with or higher than the Lifshitz energy scale.}

\begin{document}

\begin{flushright} {\footnotesize YITP-21-07, IPMU21-0006}  \end{flushright}

\maketitle
\flushbottom

\section{Introduction}
The Ho\v{r}ava-Lifshitz (HL) gravity theory \cite{Horava:2009uw} is one of the most promising candidates for a  quantum theory of gravity. In this theory, space (${\bf x}$) and time ($t$) are anisotropic and they follow the Lifshitz scaling
\begin{equation}
{\bf x} \to b {\bf x}, \ \ t \to b^z t,
\end{equation}
where $b$ is a constant and $z$ is the dynamical critical exponent. In order for the theory to be renormalizable (at least in a power-counting level\footnote{It was rigorously shown in \cite{Barvinsky:2015kil} that the projectable HL theory is renormalizable.}), $z=3$ is required~\footnote{For $z>3$ the theory is super-renormalizable at least in the sense of power-counting.}. The theory of HL gravity has advantage not only in the context of quantum gravity but also in cosmology. For example, the Hamiltonian constraint in the projectable HL gravity is not a local equation but an integrated equation, which allows the emergence of dark matter as integration constant \cite{Mukohyama:2009mz,Mukohyama:2009tp}. Also, the superluminality caused by the anisotropy of the $z=3$ HL gravity can solve the horizon problem and lead to scale-invariant cosmological perturbations \cite{Mukohyama:2009gg,Kiritsis:2009sh}. This means that the HL gravity provides an alternative to inflationary cosmology~\footnote{The flatness problem may also be addressed by the Lifshitz scaling~\cite{Kiritsis:2009sh,Bramberger:2017tid}.}. In this sense, the theory of HL gravity is well-motivated by cosmological considerations\footnote{For a review of the cosmological aspect of the HL gravity as well as the "Vainshtein screening" for the scalar graviton, see Ref. \cite{Mukohyama:2010xz}} while being one of the promising candidates for the theory of quantum gravity.

However, the Lifshitz scaling breaks the Lorentz symmetry of gravity and the Lorentz breaking is significant at short-length scales of $\lesssim 1/M_{\rm HL}$. Could any novel phenomena happen if spacetime has such a microscopic length scale caused by the anisotropy between space and time? In this paper, we will consider this interesting issue by focusing on the perturbation of a static black hole. Black hole perturbation theory demonstrates many non-trivial features, such as superradiance \cite{Teukolsky:1974yv,Brito:2015oca} and the universality of the late-time ringdown waveform \cite{Chandrasekhar:1975zza,Berti:2009kk}. Although the ringing behavior has been investigated in the framework of the HL gravity with a covariant (no-higher-derivative) scalar field \cite{Chen:2009gsa}~\footnote{Note, however, that ref.~\cite{Chen:2009gsa} considers a non-projectable HL theory without terms depending on the spatial derivatives of the lapse function and such a setup is known to be inconsistent.}, to the best of our knowledge, it has not yet been investigated how the spatial higher-derivative terms affect the ringing behavior and reflectivity of black holes.
\begin{figure}[t]
\centering
    \includegraphics[width=1\textwidth]{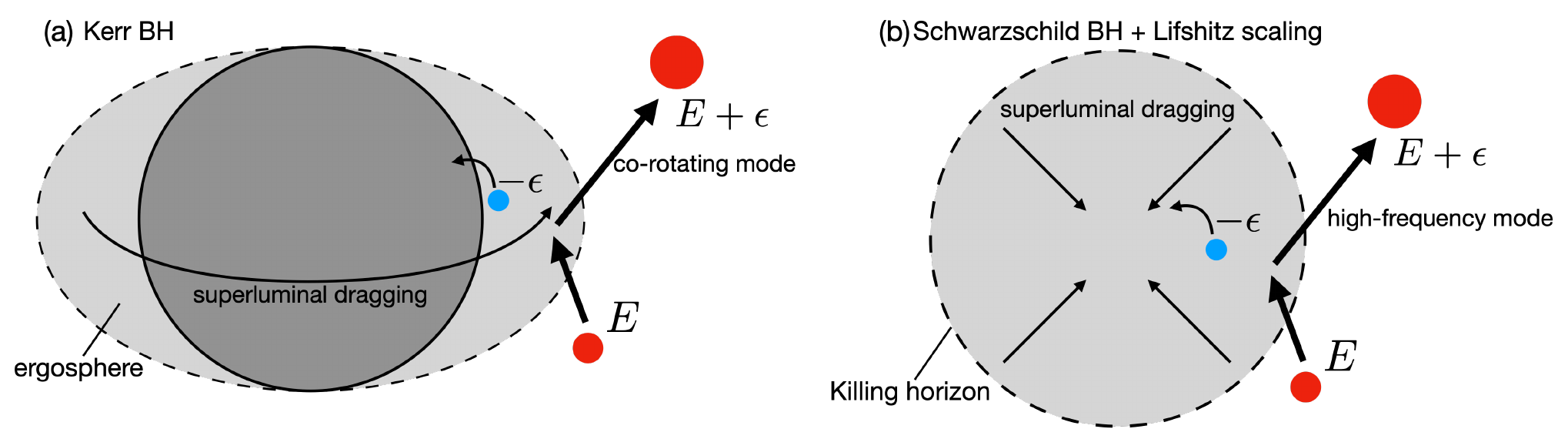}
\caption{A schematic picture showing the superradiant scattering around (a) a Kerr black hole and (b) Schwarzschild black hole with a Lifshitz field. The negativity of energy is allowed in the region where spacetime is superluminally dragged such as the ergosphere of a Kerr black hole or the interior of Killing horizon of a Schwarzschild black hole. Therefore, modes leaving from such a region can carry out additional positive energy to infinity while leaving negative energy there. This is nothing but the superradiance effect.
}
\label{super_picturte}
\end{figure}

By numerically solving the Lifshitz field equation, we find that the standard behaviour of ringing black hole, i.e. the exponential suppression and constant-frequency oscillation (see \cite{Berti:2009kk} for a review), is not guaranteed. Depending on the parameters characterizing the Lifshitz theory, we find novel features such as a power-law tail without ringing oscillations or long-lived quasinormal modes. We also compute the reflectivity of scattered Lifshitz waves by a Schwarzschild black hole and find that it exhibits the superradiance even though the black hole has no angular momentum. This is not surprising since the Killing horizon is no longer a causal boundary due to the superluminality of the Lifshitz field, and high-frequency modes could enter and leave the interior of the Killing horizon while low-frequency modes are still trapped. This is very similar to the nature of ergosphere of a Kerr black hole where co-rotating modes can enter and leave the ergosphere and the counter-rotating modes are trapped. The (superluminal) spacetime dragging inside the ergosphere allows negative energy to exist, which allows co-rotating modes to carry out some additional positive energy to infinity while leaving negative energy in the ergosphere. Similarly, negative energy can exist inside the Killing horizon of a Schwarzschild black hole, and superluminal modes could extract some positive energy out of the horizon (FIG. \ref{super_picturte}). In contrast, if the superluminal propagation is prohibited due to Lorentz invariance, the negative energy inside the horizon is causally disconnected from outside, which is why the superradiance never occurs for standard Schwarzschild black hole. We expect that the superradiance would be more significant for a smaller black hole whose Hawking temperature is higher than the Lifshitz energy scale $M_{\rm HL}$. This may drastically change the greybody factor and evaporation rate at the final stage of black hole evaporation \cite{Hawking:1974sw,Hawking:1974rv}, which might be testable via the observation of stochastic gravitational waves (GWs) since some small primordial black holes (if existed) would have evaporated at the early stage of the Universe and may have caused the sudden reheating process, which results in inducing stochastic GWs \cite{Inomata:2020lmk}. 

The superradiance is closely related to Penrose process and the latter is expected to occur in theories with spontaneous breaking of the Lorentz symmetry \cite{Eling:2007qd}. The argument is that when particles of different species interact with the ghost condensate \cite{ArkaniHamed:2003uy} and their propagation speeds are different, two apparent horizons appear with different radii and the Penrose process is made possible in the region between the two horizons. In Ref. \cite{Dubovsky:2006vk}, the apparent violation of the generalized second law (GSL) was studied in that setup, and their gedanken experiment showed that a perpetuum mobile involving a black hole and two thermal shells could be realized. However, this is not the case \cite{Mukohyama:2009um,Mukohyama:2009rk}, at least in the original ghost condensation scenario, since the accretion rate of the ghost condensate onto the black hole \cite{Mukohyama:2005rw}, which increases the black hole entropy, overwhelms the effect of the perpetuum mobile~\footnote{See also \cite{Jazayeri:2016jav} for the compatibility of the ghost condensate with the de Sitter entropy bound introduced in \cite{ArkaniHamed:2007ky} as a closely related issue.}.
In our situation, unlike the case with ghost condensate~\cite{Mukohyama:2009rk,Mukohyama:2009um}, the GSL may be violated due to the Penrose process, and so, one might wonder if it allows construction of a perpetuum mobile of the second kind. The hierarchy between two different Hawking temperatures is essential in the above gedanken experiment \cite{Dubovsky:2006vk}. On the other hand, in our situation the universal horizon\footnote{It was pointed out that the universal horizon is unstable against the perturbations \cite{Blas:2011ni}. However, it takes infinite time to form the universal horizon in the preferred frame while the evaporation time is finite. Therefore, the stability of {\textit apparent} universal horizon originating from a gravitational collapse is still open question.} is the unique horizon and the temperature associated with the universal horizon is also uniquely determined\footnote{Here we implicitly assume that all matters have the same power of momentum in their dispersion relation at high energies. Otherwise, as shown in \cite{Herrero-Valea:2020fqa}, the Hawking temperature of the universal horizon is not unique. Also, the temperature associated with the universal horizon depends on vacuum choice. According to \cite{Herrero-Valea:2020fqa}, the inconsistency between the results in \cite{Michel:2015rsa} and \cite{Berglund:2012fk} can be explained by the difference of vacuum choice. However, in either case, the Hawking temperature can be unique for the universal horizon.} \cite{Berglund:2012fk,Herrero-Valea:2020fqa}. Therefore, the perpetuum mobile would not be allowed in our case, at least in the same manner as Ref. \cite{Dubovsky:2006vk}. On a separate note, it is even clear if the violation of the GSL is problematic. For example, the Hawking-Moss transition \cite{Hawking:1981fz} also violates the GSL \cite{Oshita:2016oqn,Oshita:2017hsb,Gregory:2020cvy,Gregory:2020hia} where the cosmological horizon shrinks. Moreover, the Jarzynski equality \cite{Jarzynski_1997_1,Jarzynski_1997_2} in the non-equilibrium statistical mechanics implies that the second law of thermodynamics can be violated.

In the next section, we introduce a simplified model of the HL gravity, where the tensor perturbation is modeled by a massless scalar field $\psi$, and briefly review the appearance of a preferred frame and a universal horizon due to the extra scalar degree of freedom $\varphi$, often called Khronon. We also explain the methodology of our numerical computation. In section \ref{sec:result}, we will show our results of the black hole ringing at late time. Also, the reflectivity of scattered waves around the black hole is investigated, and it is found out that the superradiance occurs due to the Lifshitz scaling. In section \ref{sec:conclusion}, we summarize our achievements and discuss the possibility of a perpetuum mobile in our case. We will use the notation $(-,+,+,+)$ throughout the manuscript.

\section{Formalism}
We will investigate the following simplified model to see how the universal features of the ringdown and reflectivity of a static black hole are affected by the Lifshitz scaling:
\begin{equation}
L = \int d^4 x \sqrt{-g} \left[ {\cal L}_{\rm EH} + {\cal L}_{\rm SG} + {\cal L}_{\rm GW} \right], \label{khrononmetric_theo}
\end{equation}
\begin{align}
{\cal L}_{\rm EH} &\equiv \frac{1}{16 \pi G} R,\\
{\cal L}_{\rm SG} &\equiv \frac{1}{16 \pi G} \left\{ \alpha (u^{\mu} \nabla_{\mu} u_{\nu})^2 - \beta \nabla_{\mu} u^{\nu} \nabla_{\nu} u^{\mu} - \gamma (\nabla_{\mu} u^{\mu})^2 \right\},\\
{\cal L}_{\rm GW} &\equiv -\psi ({\cal F}(\Delta) + \Box) \psi,
\end{align}
where $\psi$ is a scalar field modeling the tensor perturbation, ${\cal F} (\Delta) \equiv  \Delta^3/M_{\rm HL}^4 - \nu_4 \Delta^2/M_{\rm HL}^2$, $\Delta$ is the Laplacian on the constant-$\varphi$ hypersurfaces (the definition of which will be given in subsection \ref{subsec:Lifshitz-wave-equation}), and $\nu_4$ is a constant of the order of unity. The unit normal vector $u_{\mu}$ is expressed in terms of the khronon field $\varphi$
\begin{equation}
u_{\mu} \equiv \frac{\partial_{\mu} \varphi}{\sqrt{\nabla_{\nu} \varphi \nabla^{\nu} \varphi}}.
\end{equation}
This theory models a situation where the background is given by a solution of the Einstein equations in general relativity\footnote{\label{footnote:decoupling}The contribution of the scalar-graviton to the background spacetime can be negligible because the parameters $\alpha$, $\beta$, and $\gamma$ are assumed to be much smaller than unity. Indeed, $\alpha$ and $\beta$ are required to be small by the observational constraints. On the other hand, either $|\gamma|\ll 1$ or $\gamma=\mathcal{O}(1)$ is compatible with the constraints. See \cite{Gumrukcuoglu:2017ijh} and (\ref{eqn:PPN1})-(\ref{eqn:GW170817}) below.} whereas gravitational perturbations (modeled by a scalar field $\psi$) follows the Lifshitz scaling at short-length scales. To discuss what situations can be covered  by this simple model, we will come back to this point in the discussion section. In this paper, we will consider the Schwarzschild background, whose line element is
\begin{equation}
ds^2 = -\left( 1- \frac{r_s}{r} \right) dt^2 + \left( 1- \frac{r_s}{r} \right)^{-1} dr^2 +r^2 d\Omega_2^2,
\end{equation}
and investigate how the Lifshitz scaling affects the scattering process around the static black hole.

\subsection{Preferred frame and the universal horizon}
The theory in (\ref{khrononmetric_theo}) has a preferred direction given by $u_{\mu}$, which stems from the preferred frame ($\varphi=const.$) one should respect. We here briefly review the preferred frame and the universal horizon based on Ref. \cite{Blas:2011ni}. The dynamics of the khronon field induces the preferred frame. The khronon field equation in the Schwarzschild background is given by \cite{Blas:2011ni}
\begin{equation}
\frac{\partial_{\xi}^2 U}{U} - c_{\chi}^2 \frac{\partial_{\xi}^2 V}{V} + \frac{2 c_{\chi}^2}{\xi^2}=0,\label{khronon_eq2_1}
\end{equation}
where $c_{\chi} \equiv \sqrt{(\beta+\gamma)/\alpha}$ and
\begin{equation}
U \equiv u_t, \ \ V\equiv u^r, \ \ \xi \equiv \frac{r_s}{r} = \frac{1}{r}.
\end{equation}
Here, we have set $r_s = 1$. The unit normal vector $u^{\mu}$ satisfies
\begin{equation}
(u_t)^2 - (u^{r})^2 = 1-\xi,
\label{khronon_eq2}
\end{equation}
and so the relation between $U$ and $V$ is given by
\begin{equation}
U^2 -V^2 = 1-\xi. \label{khronon_eq2_2}
\end{equation}
Choosing the branch with in-going $u^{\mu}$ (i.e. $V = u^r < 0)$ and thus plugging $V = -\sqrt{U^2 -1 + \xi}$ into (\ref{khronon_eq2_1}), one obtains
\begin{equation}
\partial_{\xi}^2 U + \frac{c_{\chi}^2 U}{U^2 (1-c_{\chi}^2) - 1 + \xi} \left[ - (\partial_{\xi} U)^2 + \frac{(U \partial_{\xi} U + 1/2)^2}{U^2 -1 + \xi} + \frac{2 (U^2 - 1+\xi)}{\xi^2} \right] = 0.
\label{univresal_eq1}
\end{equation}
One can also rewrite the background metric as
\begin{equation}
ds^2 = -(1-\xi) d\tau^2 - 2 V d\tau dr^{\ast} + dr^{\ast} {}^2 + r^2 d\Omega_2^2,
\label{tortoise_at_univ}
\end{equation}
where $d\tau = dt - \frac{V}{1-\xi} dr^{\ast}$ and $dr^{\ast} = dr/U$. Note that $r^*$ differs from the standard definition of tortoise coordinate in general relativistic black holes, as $r^* \to -\infty$ refers to the universal (not Killing) horizon where $U \to 0$.

When $U=1$, the metric (\ref{tortoise_at_univ}) reduces to the one in the  Gullstrand-Painlev\'{e} coordinates
\begin{equation}
ds^2 = -d\tau^2 + (dr+\sqrt{\xi}d\tau)^2 +r^2 d\Omega_2^2.
\end{equation}
The sound horizon appears at $\xi = \xi_c$ that satisfies
\begin{equation}
U^2(\xi_c) (1-c_{\chi}^2) = 1- \xi_c.
\end{equation}
In order for the second term in (\ref{univresal_eq1}) to be regular, one has to impose
\begin{equation}
\partial_{\xi} U (\xi_c) = \frac{1}{2 (1-c_{\chi}^2) U(\xi_c)} \left[ -1 + c_{\chi} \sqrt{1- \frac{8 c_{\chi}^2 (1- c_{\chi}^2) U^4 (\xi_c)}{\xi_c^2}} \right].
\label{UH_eq}
\end{equation}
Now imposing the boundary condition of $U(0) = 1$ and using the shooting method, one can numerically solve (\ref{univresal_eq1}). When $c_{\chi} \to 0$ or $c_{\chi} \to \infty$, (\ref{univresal_eq1}) has analytic solutions
\begin{align}
U (\xi) =
\begin{cases}
1-\frac{\xi}{2} & \text{($c_{\chi} \to 0$)},\\
\sqrt{1-\xi + \frac{27}{16^2} \xi^4} & \text{($c_{\chi} \to \infty$)}.
\end{cases}
\end{align}
In the next section, we will investigate the perturbations of the Lifshitz scalar field in the both limits: $c_{\chi} \to \infty$ and $c_{\chi} \to 0$. The two limits can be compatible with the observational and theoretical constraints on the parameters obtained in Ref. \cite{Gumrukcuoglu:2017ijh}. Most of the constraints are satisfied for $\alpha$, $\beta$, $\gamma \ll 1$~\footnote{The vacuum Cherenkov constraint from the scalar graviton was not considered in \cite{Gumrukcuoglu:2017ijh} and this treatment seems consistent with the decoupling limit implied by $\alpha$, $\beta$, $\gamma \ll 1$. This may not have been the case if $\gamma =\mathcal{O}(1)$ (see footnote~\ref{footnote:decoupling}).}. The non-trivial constraints are the constraints on the parametrized post-Newtonian (ppN) parameters quantifying preferred-frame effects, which translate to \cite{Gumrukcuoglu:2017ijh}
\begin{align}
\left| \frac{4 (\alpha - 2 \beta)}{1-\beta} \right| \lesssim 10^{-4}, \label{eqn:PPN1}\\
\left| \left( \frac{\alpha -2\beta}{2 - \alpha} \right) \left(1- \frac{(\alpha -2\beta) (1+ \beta + 2\gamma)}{(1-\beta) (\beta + \gamma)} \right) \right| \lesssim 10^{-7}.
\end{align}
Also, the observation of gravitational waves emitted from the event GW170817 with the gamma ray emission put a stringent constraint on $\beta$
\begin{equation}
|\beta| \lesssim 10^{-15}. \label{eqn:GW170817}
\end{equation}
Assuming $\beta = 0$, for example, the ppN constraints become $4 |\alpha| \lesssim 10^{-4}$ and $(|\alpha|/2) |1- c_{\chi}^{-2}| \lesssim 10^{-7}$. In the two cases, $c_{\chi} \gg 1$ and $c_{\chi} \ll 1$, the latter constraint reduces respectively to
\begin{align}
|\alpha| \lesssim 2 \times 10^{-7} & \ \ (c_{\chi} \gg 1),\label{ccgg}\\
c_{\chi}^2 \gtrsim |\alpha| \times 5 \times 10^{6} & \ \ (c_{\chi} \ll 1)\label{ccll},
\end{align}
Therefore, it turns out that the limit of $c_{\chi} \to \infty$ is compatible with the constraints once taking a small value of $\alpha$ so that (\ref{ccgg}) is satisfied. The other limit $c_{\chi} \to 0$ may be also compatible\footnote{But note that $c_{\chi} \to 0$ limit may lead to development of non-perturbative behavior due to caustic formation.} with them when taking the infinitesimal value of $\alpha$.

\subsection{Lifshitz wave equation} \label{subsec:Lifshitz-wave-equation}
We will investigate the dynamics of incoming scalar waves around a static black hole based on the following wave equation:
\begin{equation}
\left[ \frac{\Delta^3}{M_{\rm HL}^4} - \nu_4 \frac{\Delta^2}{M_{\rm HL}^2} + \Box \right] \psi(t,r, \theta, \phi) = 0,
\label{eom_scalar}
\end{equation}
where $\Delta \equiv D_a D^a$ and $D_a$ is the covariant derivative on the khronon surfaces, $M_{\rm HL}$ is the Lifshitz energy scale, and $\nu_4$ is a dimensionless parameter. In order to obtain the explicit form of $D_a$, let us decompose the metric as
\begin{equation}
ds^2 = -N^2 d\tau^2 + h_{ij} (dx^i + N^i d\tau) (dx^j + N^j d \tau).
\end{equation}
Comparing it with (\ref{tortoise_at_univ}), one can read
\begin{equation}
N^2 = (1- \xi) + V^2, \
N^{r} = -V, \
h_{ij} = \text{diag} (1, \ r^2, \ r^2 \sin^2 \theta).
\end{equation}
The definition of $D_i v^j$ is
\begin{equation}
D_i v^j = \partial_i v^j + {}^{(3)} \Gamma^j_{ik} v^k,
\end{equation}
and ${}^{(3)} \Gamma^j_{ik}$ is the Levi-Civita connections w.r.t. $h_{ij}$:
\begin{equation}
{}^{(3)} \Gamma^j_{ik} = \frac{1}{2} h^{jl} (\partial_i h_{lk} + \partial_k h_{il} - \partial_l h_{ik}).
\end{equation}
The explicit form of the Levi-Civita connections in three-space is presented in Appendix \ref{app_LCCs_LAPs}. Therefore, the Laplacian on the khronon surface is
\begin{align}
\Delta \psi = D_i D^i \psi = D_i \partial^i \psi = \partial_i \partial^i \psi + {}^{(3)} \Gamma^i_{ik} \partial^k \psi = \partial_i h^{ij} \partial_j \psi + {}^{(3)} \Gamma^i_{ik} \partial^k \psi.
\end{align}
Let us explicitly write down the d'Alembertian by using the metric (\ref{tortoise_at_univ}). The covariant part of the equation of motion is
\begin{equation}
\Box \psi = g^{\mu \nu} \partial_{\mu} \partial_{\nu} \psi - g^{\mu \nu} \Gamma^{\alpha}_{\mu \nu} \partial_{\alpha} \psi = 0,
\end{equation}
and the inverse metric $g^{\mu \nu}$ is
\begin{equation}
g^{\mu \nu} =
\begin{pmatrix}
-1/U^2 & -V/U^2 & 0 & 0\\
-V/U^2 & (1-1/r)/U^2 & 0 &0\\
0 & 0 & 1/r^2 & 0\\
0 & 0 & 0 & 1/(r^2 \sin^2 \theta)
\end{pmatrix}.
\end{equation}
Then the Lorentz breaking equation of motion (\ref{eom_scalar}) can be explicitly written as
\begin{align}
\begin{split}
&U^2 \left( -\frac{\Delta^3}{M_{\rm HL}^4} + \kappa \frac{\Delta^2}{M_{\rm HL}^2} \right) \Psi\\
&+ \left[ \partial_{\tau}^2 - \left( 1-\frac{1}{r} \right) \partial_{r^{\ast}}^2 + 2 V \partial_{\tau} \partial_{r^{\ast}} + U^2 \frac{\ell (\ell + 1)}{r^2} \right] \Psi\\
&+\frac{1}{U^2} \left[ V' \left( 1- \frac{1}{r} \right) - \frac{UV}{2r^2} \right] \left( \partial_{\tau} + V \partial_{r^{\ast}} \right) \Psi\\
&- \frac{2U}{r} \left[ 1-\frac{3}{4 r} \right] \partial_{r^{\ast}} \Psi + \frac{2UV}{r} \partial_{\tau} \Psi = 0,
\end{split}
\label{eom_final}
\end{align}
where
\begin{align}
\psi &= \Psi (t,r) Y_{\ell m}(\theta, \phi),\\
\Delta &= \partial_{r^{\ast}}^2 + \frac{2 U}{r} \partial_{r^{\ast}} - \frac{\ell (\ell + 1)}{r^2}.
\label{lap}
\end{align}
The explicit form of the quadratic and cubic Laplacians are shown in Appendix \ref{app_LCCs_LAPs}.

\subsection{Numerical methodology}\label{sec:num}
We numerically solve the wave equation (\ref{eom_final}) with the 4th-order Runge-Kutta method. First, we decompose it into two first-order differential equations with respect to $\tau$
\begin{align}
\frac{d \Psi}{d \tau} &= \Pi (\tau, r),\\
\frac{d \Pi}{d \tau} &= -U^2 \left( -\frac{\Delta^3}{M^4} + \kappa \frac{\Delta^2}{M^2} \right) \Psi\\
&- \left[- \left( 1-\frac{1}{r} \right) \partial_{r^{\ast}}^2 + U^2 \frac{\ell (\ell + 1)}{r^2} \right] \Psi -2V \partial_{r^{\ast}} \Pi\\
&-\frac{1}{U^2} \left[ V' \left( 1- \frac{1}{r} \right) - \frac{UV}{2r^2} \right] \left( \Pi + V \partial_{r^{\ast}} \Psi \right)\\
&+ \frac{2U}{r} \left[ 1-\frac{3}{4 r} \right] \partial_{r^{\ast}} \Psi - \frac{2UV}{r} \Pi,
\end{align}
where we introduced a new function $\Pi \equiv d \Psi/ d\tau$, and we should solve $\Psi (\tau , r^{\ast})$ and $\Pi (\tau, r^{\ast})$ simultaneously. We compute the spatial derivative terms $\partial_{r^{\ast}}^n X$ ($X$ is $\Psi$ or $\Pi$) with the \textit{Mathematica}'s function {\texttt{NDSolve`FiniteDifferenceDerivative}}. In this function, the derivatives at the spatial boundaries are calculated with one-sided formulas.

In the following, we use two parameters $\Xi_4$ and $\Xi_6$ defined as
\begin{equation}
\Xi_4 \equiv \frac{\nu_4}{M_{\rm HL}^2r_s^2}, \ \Xi_6 \equiv \frac{1}{M_{\rm HL}^4 r_s^4}.
\label{XI_4and6}
\end{equation}
In this definition, the dispersion relation at infinity becomes
\begin{equation} \label{eqn:dispersionrelation}
\omega^2=\Xi_6 k^6 +\Xi_4 k^4 + k^2,
\end{equation}
where $\omega$ and $k$ is a frequency and wavenumber with respect to $\tau$ and $r^{\ast}$, respectively. The ratio of the time step $\Delta \tau$ to the spatial step $\Delta r^{\ast}$ ($\lambda \equiv \Delta \tau/ \Delta r^{\ast}$) should be fixed with a value smaller than unity, in order to satisfy an approximate Courant condition for the superluminal modes that propagate outside the light cone. In Appendix \ref{app_conv}, we simulate the wave propagation with $\lambda = 0.2$, $0.1$, and $0.05$, and found out that the results converge for $\lambda \lesssim 0.1$ when $|\Xi_4|\lesssim 0.1$ and $\Xi_6 \lesssim 0.01$, provided that the typical value of the wavenumber $k$ is of order unity. Therefore, we will use $\lambda = 0.1$ in the following analysis. However, note that a computation involving stronger superluminal modes (e.g., $\Xi_4 \gg 0.1$ or $\Xi_6 \gg 0.01$ for typical wavenumber of order unity) may need a smaller value of $\lambda$. In the Appendix \ref{app_conv}, we also confirm the consistency of our simulations with the known results (i.e., the fundamental quasinormal mode and reflectivity) of the Lorentz invariant case ($\Xi_4=\Xi_6=0$). We control numerical high-frequency unstable modes with the Kreiss-Oliger dissipation\cite{kreiss1973methods} of an amplitude of $1/16$.

As an initial configuration of the Lifshitz scalar field, we assume wave packet centered at $r= r_w$ with the form of
\begin{align}
\Psi (\tau=0, r^{\ast}) &= \exp{\left[ -\frac{(r^{\ast}-r_w)^2}{s^2} \right]} \cos{\tilde{\omega} r^{\ast}}, \label{ini_1}\\
\dot{\Psi} (\tau=0, r^{\ast}) &= -\left( \frac{2 (r^{\ast} -r_w)}{s^2} \cos \tilde{\omega} r^{\ast} + \tilde{\omega} \sin \tilde{\omega} r^{\ast} \right) \exp{\left[ -\frac{(r^{\ast}-r_w)^2}{s^2} \right]}.
\label{ini_2}
\end{align}
We use $s=2$, $\tilde{\omega} = 1$ throughout the analysis, meaning that the typical value of the wavenumber $k$ in the dispersion relation (\ref{eqn:dispersionrelation}) is of order unity. Although this is purely ingoing waves at infinity for $\Xi_4=\Xi_6=0$, this leads to partial outgoing modes when the dispersion relation is modified due to the Lifshitz scaling. 
There are two types of modified dispersion relation: $\Xi_4 > 0$ and $\Xi_4 < 0$. In analogy to superfluid perturbations, we call the latter case a {\it roton} dispersion relation, as it could lead to backward propagation in a range of frequency/distance from the black hole. The sign of the sixth-derivative terms should be positive in order to guarantee the renormalizability and the UV stability. 

\section{Results}
\label{sec:result}
With the initial condition (\ref{ini_1}) and (\ref{ini_2}), we numerically solve for $\Psi (t, r^{\ast})$, following the prescription outlined in Section \ref{sec:num}. Some snapshots of these solutions are shown in Appendix \ref{app_snap}. In this section, we present and analyze the detailed results for the ringdown and reflectivity of the black hole.

\subsection{Ringdowns, Rotons, and Long-lived modes}
\begin{figure}[t]
\centering
    \includegraphics[width=1\textwidth]{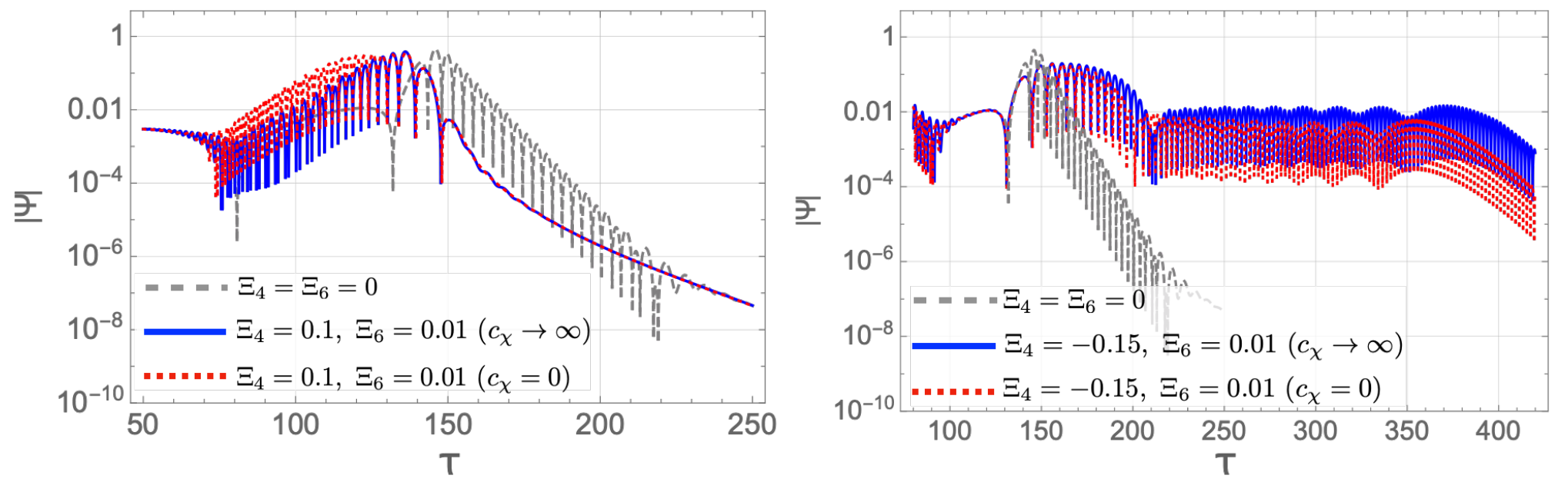}
\caption{Time domain functions of $|\Psi|$ with $r^{\ast}_o = 60$ and $r^{\ast}_w =80$. The cases of Lifshitz scaling with $\Xi_4=0.1$ (left) and $\Xi_4=-0.15$ (right) are shown. The Lorentz invariant case (gray-dashed) is also shown for comparison.
}
\label{time_domain}
\end{figure}

\begin{figure}[t]
\centering
    \includegraphics[width=1\textwidth]{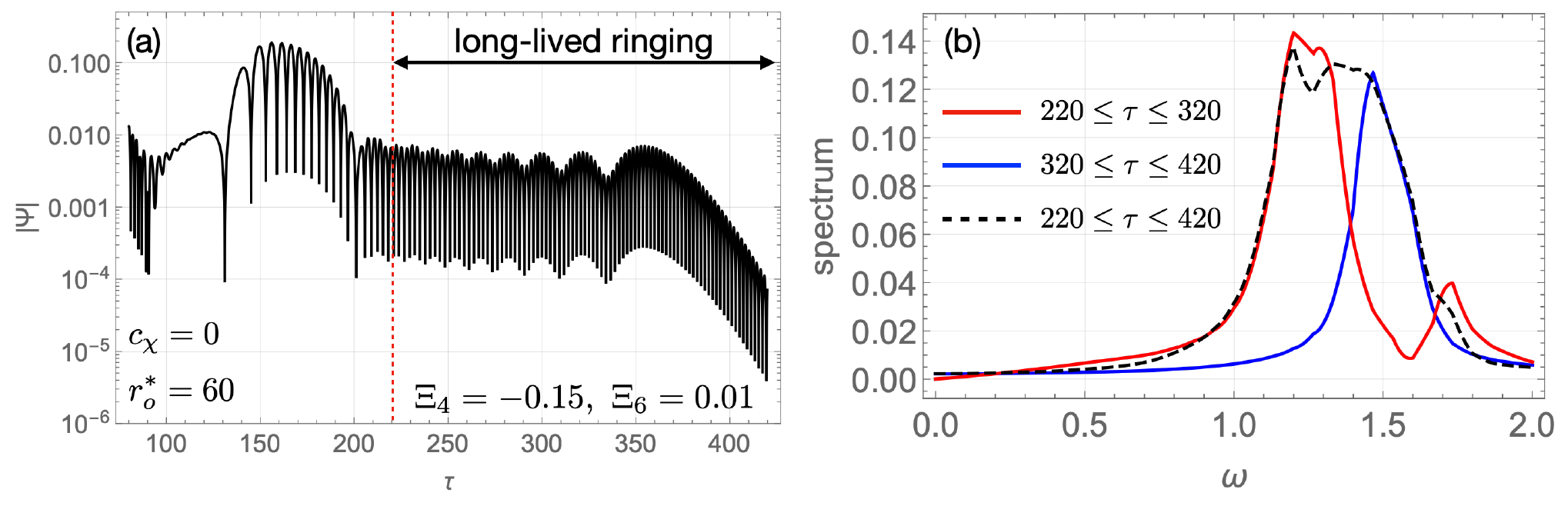}
\caption{(a) The time domain function for $\Xi_4 =-0.15$ and $\Xi_6 =0.01$. (b) The absolute value of the spectrum for the time domain function in the range of $220 \leq \tau \leq 320$ (red), $320 \leq \tau \leq 420$ (blue), and $220 \leq \tau \leq 420$ (black-dashed).
}
\label{spectrum}
\end{figure}

\begin{figure}[t]
\centering
    \includegraphics[width=0.7\textwidth]{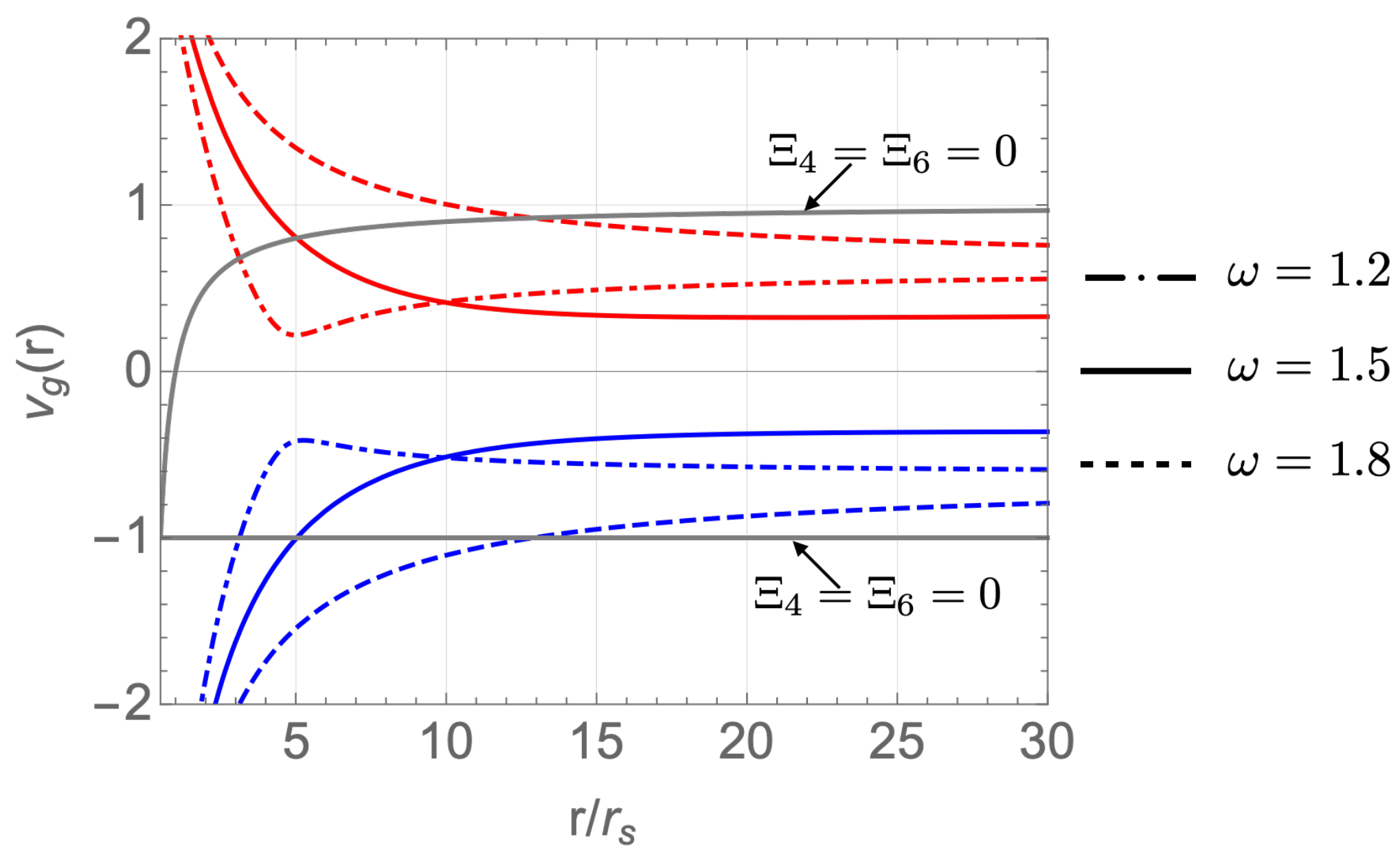}
\caption{The trajectories in the phase space obtained from (\ref{dr_simplified}). We use the same parameters as in FIG. \ref{spectrum}. The red and blue lines represent the outgoing and ingoing modes, respectively. The gray lines show the trajectories for the Lorentz invariant case ($\Xi_4=\Xi_6=0$).
}
\label{group_velocity}
\end{figure}

In order to see the late-time behavior of $\psi$ with the Lifshitz scaling, we calculate $\Psi (t, r^{\ast}_o)$, where $r^{\ast}_o$ is the position of an observer. The results are shown in FIG. \ref{time_domain}.
For the case of $\Xi_4 > 0$ (left panel of FIG. \ref{time_domain}), the quasinormal ringing are suppressed and the power-law tail appears earlier than the Lorentz invariant case (grey line in FIG. \ref{time_domain}). Also, high-frequency modes appear earlier than the power-law tail. The form of the tail is universal and independent of the parameters of Lifshitz scaling. This is consistent with the fact that the late-time tail is caused by back-scattering off the background curvature, and therefore, its power depends only on the asymptotic background spacetime \cite{Berti:2009kk}. On the other hand, the ringdown lasts longer for the roton dispersion relation, $\Xi_4 < 0$. In this cases, as we show below, the group velocity $v_g\equiv d\omega/dk$ is suppressed (enhanced) at the intermediate (high) frequencies in this case, and therefore wavepackets are dispersed significantly in space/time. We believe this is the origin of the long-lived ringing at late time. We computed the spectrum of the long-lived modes for $(\Xi_4, \Xi_6) = (-0.15,0.01)$ and found that the dominant modes are around $\omega \approx 1.5$ (black dashed line in FIG. \ref{spectrum}-b), which can be explained using the following simple analytic estimate: To see this, let us simplify the modified dispersion relation as\footnote{The first two terms in (\ref{dr_simplified}) represent the inward frame dragging and the last term gives the Lifshitz scaling at high frequencies. We assume that the simplified dispersion relation captures the essence of wave propagation with the frame dragging and Lifshitz scaling. We also confirmed that the numerical result (FIG. \ref{spectrum}) is well consistent with the analysis based on the simplified dispersion relation (FIG. \ref{group_velocity}).}
\begin{equation}
\omega^2 = \left(1-\frac{1}{r} \right) k^2 + 2 V(r) k \omega +U^2(r)\left(\Xi_4 k^4+\Xi_6 k^6 \right).
\label{dr_simplified}
\end{equation}
For the case of $c_{\chi} = 0$, the group velocities of ingoing and outgoing modes are
\begin{align}
v_g(r,k) = - \frac{1}{2r}- \left( 1-\frac{1}{2r} \right) \sqrt{1+2 \Xi_4 k^2 + 3 \Xi_6 k^4} <0,\\
v_g(r,k) = - \frac{1}{2r}+ \left( 1-\frac{1}{2r} \right) \sqrt{1+2 \Xi_4 k^2 + 3 \Xi_6 k^4} >0,
\end{align}
respectively, and the position $r = r(\omega,k)$ is obtained by solving (\ref{dr_simplified})
\begin{equation}
r(\omega,k) = \frac{\omega_{\rm flat}(k) +k}{2(\omega_{\rm flat}(k)-\omega)},
\end{equation}
with $\omega_{\rm flat}(k) \equiv \sqrt{k^2+ \Xi_4 k^4+ \Xi_6 k^6}$.
The incoming and outgoing trajectories in the phase diagram ($v_g-r$ plane) are shown in FIG. \ref{group_velocity}. Note that the ingoing and outgoing trajectories are separated and do not describe the reflection at the angular momentum potential as the modified dispersion relation shown in (\ref{dr_simplified}) does not include the potential term. Therefore, the trajectories around $r\lesssim \ell/\omega$ shown in FIG. \ref{group_velocity} is not reliable. As shown in FIG. \ref{group_velocity}, the group velocity is indeed suppressed for intermediate frequencies around $\omega \simeq 1.5$. This is consistent with the fact that the late-time ringing within $320 \leq \tau \leq 420$ are dominated by the modes of $\omega \sim 1.5$ (see the blue line in FIG. \ref{spectrum}-(b)). On the other hand, the neighbouring modes $\omega \sim 1.2$ and $\omega \sim 1.8$ get out earlier (see the red line in FIG. \ref{spectrum}-(b)), which is also consistent with the analytically obtained trajectories in the phase space (FIG. \ref{group_velocity}) as the group velocities for $\omega=1.2$ and $\omega=1.8$ are higher than that for $\omega = 1.5$.

\subsection{Reflectivity and Superradiance}
\begin{figure}[t]
\centering
    \includegraphics[width=1\textwidth]{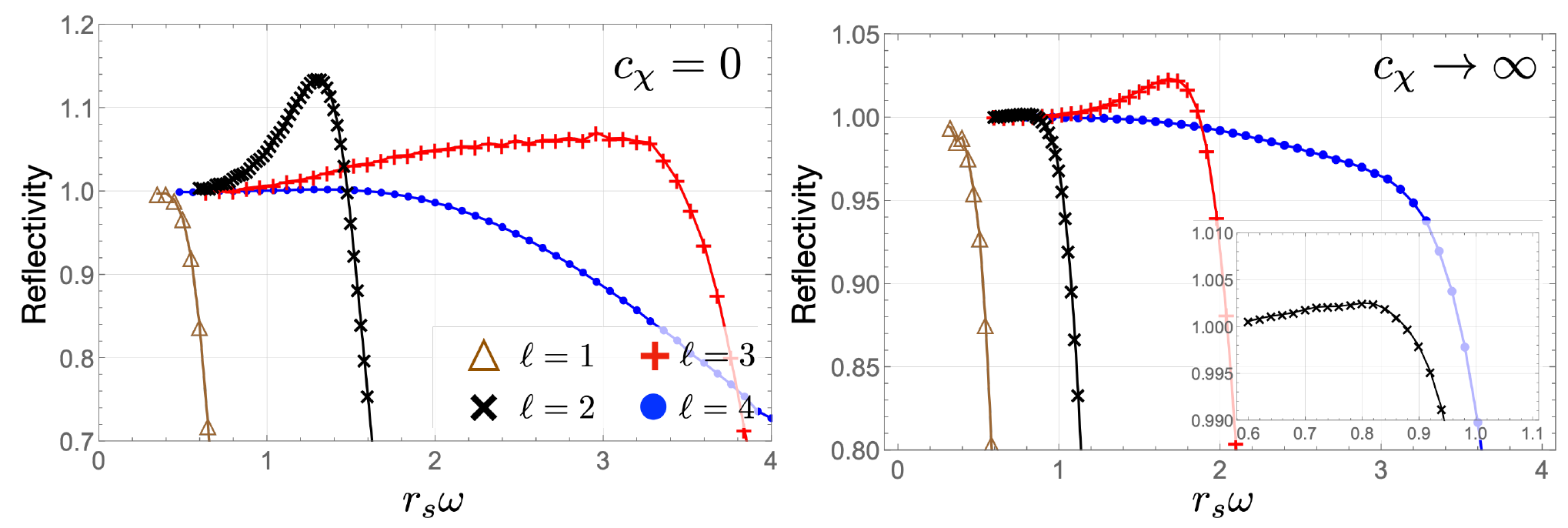}
\caption{The reflectivity of the static black hole with the Lifshitz scalar field of $\Xi_4 = 0.1$ and $\Xi_6 = 0.01$. The left and right panels show the frequency-dependence of reflectivity for $c_{\chi} = 0$ and $c_{\chi} \to \infty$, respectively. For $\ell = 2$ and $3$, the reflectivity exceeds unity, which means that the non-spinning black hole exhibits the superradiance effect. 
}
\label{reflection}
\end{figure}

\begin{figure}[t]
\centering
    \includegraphics[width=1\textwidth]{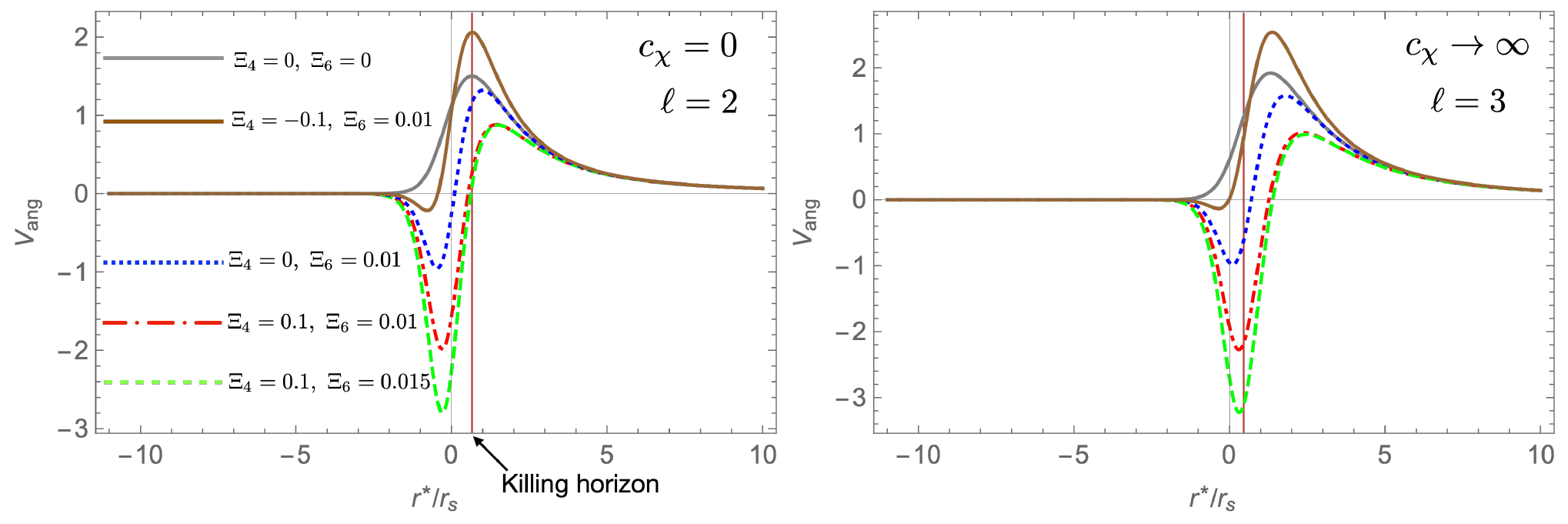}
\caption{The angular momentum potentials for the various parameters. The negative energy region inside the potential barrier is deeper for larger values of $\Xi_4(>0)$ and $\Xi_6$.
}
\label{pot}
\end{figure}

As the next exercise, we numerically calculate the reflectivity of a black hole with the (non-roton) Lifshitz scaling of $\Xi_4 > 0$. Here, we implement the Fourier transformation for the ingoing and outgoing wavepackets, measured by the observer at $r^{\ast} = r^{\ast}_{o}$, and calculate the reflectivity defined by the absolute value of the ratio between the ingoing and outgoing Fourier coefficients. The result (FIG. \ref{reflection}) shows that the superradiance (i.e. Reflectivity larger than unity) occurs for $\Xi_4 > 0$. One might wonder why the superradiance occurs even though the black hole has no angular momentum. In our situation, the superluminal propagation is allowed due to the Lifshitz scaling, and the superluminal modes can enter and leave the interior of the Killing horizon where negative energy can exist as in the ergosphere of a Kerr black hole. Therefore, superluminal modes of the Lifshitz scalar can access the interior to carry out additional positive energy to infinity while leaving the negative energy inside the Killing horizon. One can also understand the superradiance effect due to the Lifshitz scaling from the negativity of the angular momentum potential term. Let us show how the potential term is modified due to the Lorentz breaking terms ${\cal F} (\Delta)$. We here define the potential term as the term which does not involve the derivative in the wave equation. Hence the modified potential term $V_{\rm ang}(r)$ is
\begin{equation}
V_{\rm ang}(r) = U^2 \left[\frac{\ell (\ell + 1)}{r^2} + \Xi_4 D + \Xi_6 \left(D''+ \frac{2U}{r} D' - \frac{\ell (\ell + 1)}{r^2} D \right) \right],
\end{equation}
where the definition of $D=D(r)$ is given in (\ref{def_D}).
In FIG. \ref{pot}, we plot the potential term including the Lorentz breaking effect and one can see that the negative energy region locally appears inside the potential barrier, which can lead to superradiance. Therefore, we conclude that Lifshitz scaling could lead to superradiant scattering, even without angular momentum of the background black hole.

Even if the energy scale of the Lifshitz scaling $M_{\rm HL}$ is higher than the typical frequency of the ringing black hole, it eventually reaches $M_{\rm HL}$ due to the evaporation of (an isolated) black hole. At the stage where the Hawking temperature is comparable with $M_{\rm HL}$, the greybody factor would be drastically modified due to the superradiance effect. Therefore, our result implies that the final stage of the black hole evaporation can be drastically different from the standard picture, provided that the Lifshitz scaling is ubiquitous at high energy scales.
\begin{figure}[t]
\centering
    \includegraphics[width=0.65\textwidth]{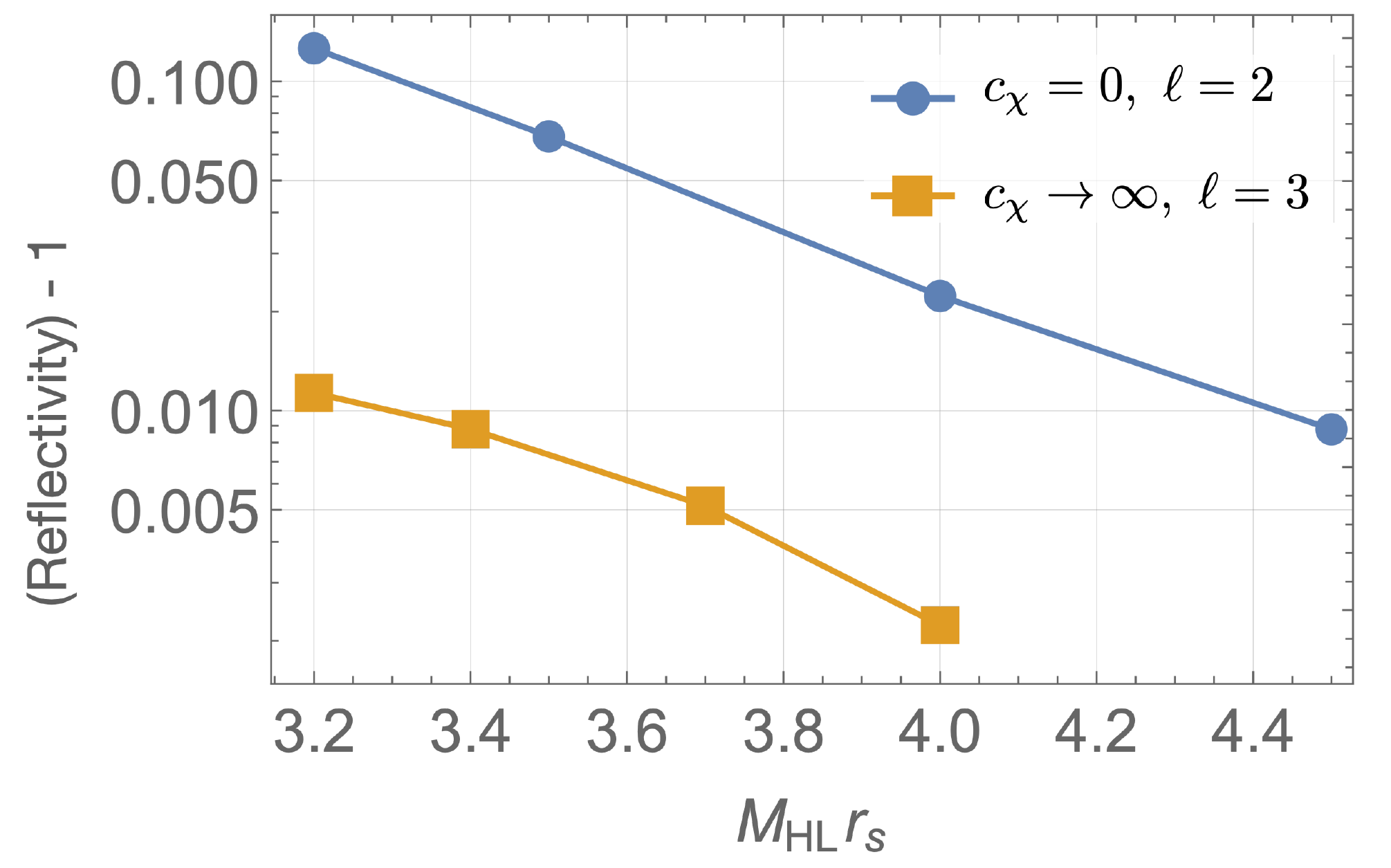}
\caption{The maximum values of the amplification factor (=(reflectivity) $-1$) for $\nu_4 = 1$. The superradiance is observed for smaller values of $r_s$.
}
\label{Hawking_rad}
\end{figure}

We compute the maximum values of the reflectivity as a function of the mass of black hole. Note that the non-dimensional parameters $\Xi_4$ and $\Xi_6$ increase as the black hole shrinks and $r_s$ becomes smaller (see Eq. (\ref{XI_4and6})). In FIG. \ref{Hawking_rad}, we show the $r_s$-dependence of the maximum value of reflectivity for $\nu_4=1$. It is challenging to extend our computation to the case of $M_{\rm HL} r_s \ll 3$ ($\Xi_6 \gg 0.01$) since higher-frequency (highly superluminal) modes are involved and a smaller value of $\lambda$ ($=$ time step/ spatial grid)  is required to numerically resolve those highly superluminal modes. Nevertheless, we expect that the trend would continue even for $M_{\rm HL} r_s \ll 3$ because the negative-energy region inside the angular momentum potential becomes deeper for a larger value of $\Xi_6$ (smaller value of black hole mass) as is shown in FIG. \ref{pot}. On the other hand, the negativity of $\nu_4$ (i.e. the roton dispersion relation) makes the negative-energy region inside the potential small, which results in quenching the superradiance. 


\section{Discussions and summary}
\label{sec:conclusion}
In this paper, we have investigated the effect of the Lifshitz scaling on the late-time ringing and reflectivity of a Schwarzschild black hole with the simplified model (\ref{khrononmetric_theo}). 
We have considered a situation where the background is given by a Schwarzschild solution whereas gravitational perturbations (modeled by a scalar field $\psi$) follows the Lifshitz scaling at short-length scales. Such a situation can be realized, for example, by considering the following minimal theory of the HL gravity with static and spherical symmetry of background
\begin{equation}
L = \int d^4 x \sqrt{-g} \left[ \frac{2}{\kappa^2} \left(K^{ij} K_{ij} - \lambda K^2 + {\cal R} \right) + \frac{\kappa^2}{2 w^4} C^{ij} C_{ij} \right],
\label{cotton_square}
\end{equation}
where $K_{ij}$ is the extrinsic curvature, $C_{ij}$ is the Cotton tensor, ${\cal R}$ is the three-dimensional Ricci scalar, $K \equiv \text{tr}[K_{\ij}]$, and $\kappa$, $\lambda$, $w$ are constants. The first three terms reduce to ${\cal L}_{\rm EH}$ and ${\cal L}_{\rm SG}$ with $c_{\chi} \to \infty$, and the quadratic action for tensorial gravitational waves is modeled by the Lorentz invariant part of ${\cal L}_{\rm GW}$ of our simplified model (\ref{khrononmetric_theo}). The second term including the Cotton tensor leads to the Lorentz breaking terms corresponding to ${\cal F} (\Delta)$ in ${\cal L}_{\rm GW}$. Let us note that if the scalar-graviton becomes dynamical under the renormalization group flow beyond some energy scale, $M_S$, then the action (\ref{cotton_square}) should contain other higher-order derivative terms (with respect to scalar-graviton) for the theory to be renormalizable. Therefore, the minimal theory (\ref{cotton_square}) can be a low-energy effective theory of quantum gravity, provided that the degree of freedom of the scalar-gravition can be traced out up to the intermediate energy scales $\sim M_{\rm HL} \ll M_S$. Our work would be applicable not only to some specific situations in the  HL gravity but also to the scattering problem of a black hole in other higher-derivative gravity theories. For example, the consistent theory of $D \to 4$ Einstein-Gauss-Bonnet gravity \cite{Aoki:2020iwm}, that amends ambiguities and fatal problems in the proposal of \cite{Glavan:2019inb}, leads to spatial higher-derivative terms in the dispersion relation .

We found out that the black hole ringing at late time disappears when the quartic derivative term is dominant with $\Xi_4 > 0$. On the other hand, the black hole ringing exhibits long-lived modulation when $\Xi_4 < 0$. We also showed that the Lifshitz waves scattered around a static black hole exhibits superradiance. The superradiance is stronger for a smaller black hole as its quasinormal frequency becomes comparable with or higher than $M_{\rm HL}$. This superradiance may significantly affect the evaporation process of a primordial black hole since it would change their greybody factor. If the energy flux of Hawking radiation is enhanced at the final stage of black hole evaporation, it could cause stronger reheating than expected before and may induce amplified stochastic gravitational waves \cite{Inomata:2020lmk} that could be observable with the future gravitational-wave detectors such as DECIGO \cite{Seto:2001qf}, BBO \cite{BBO}, and LISA \cite{2017arXiv170200786A}.

The Lifshitz scaling leads to the modifications to dispersion relation. The coefficients of the modifications have been constrained by the observations of gravitational wave by the LIGO and Virgo collaboration \cite{Abbott:2020jks}. Based on the latest observational constraint \cite{Abbott:2020jks}, the Lifshitz scaling is less important at least for the typical frequency (quasinormal frequency) of a black hole with $M\gg 10^{-8} M_{\odot}$. Therefore, the novel phenomena investigated here, at least for $\nu_4>0$, could be important only for asteroid-mass or smaller primordial black holes. For $\nu_4<0$, one may imagine high energy excitations (e.g., ultra high energy cosmic rays) that could excite long-lived roton modes, even in the vicinity of black hole horizons. 

\acknowledgments

We thank Sergey Sibiryakov for the feedback on a draft of the manuscript. The work of NO was supported in part by the JSPS Overseas 
Research Fellowships and by the Perimeter Institute for 
Theoretical Physics. 
The work of SM was supported in part by Japan
Society for the Promotion of Science Grants-in-Aid for Scientific
Research No.~17H02890, No.~17H06359, and by World Premier
International Research Center Initiative, MEXT, Japan.
Research at Perimeter Institute is supported by 
the Government of Canada through the Department of Innovation, 
Science and Economic Development Canada and by the Province of 
Ontario through the Ministry of Research, Innovation and Science.

\appendix
\section{Levi-Civita connections, quadratic, and cubic Laplacians}
\label{app_LCCs_LAPs}
In this appendix, we show explicit forms of Levi-Civita connections and quadratic/cubic Laplacians that appear in the radial wave equation (\ref{eom_final}). The non-zero Levi-Civita connections in three-space are
\begin{align}
\begin{split}
&{}^{(3)}\Gamma^{r^{\ast}}_{\theta \theta} = -r U, \
{}^{(3)}\Gamma^{r^{\ast}}_{\phi \phi} = -r U \sin^2 \theta, \
{}^{(3)}\Gamma^{\theta}_{r^{\ast} \theta} = U/r, \
{}^{(3)}\Gamma^{\theta}_{\theta r^{\ast}} = U/r,\\
&{}^{(3)}\Gamma^{\theta}_{\phi \phi} = -\cos \theta \sin \theta, \
{}^{(3)}\Gamma^{\phi}_{r^{\ast} \phi} = U/r, \
{}^{(3)}\Gamma^{\phi}_{\theta \phi} = \cot \theta, \
{}^{(3)}\Gamma^{\phi}_{\phi r^{\ast}} = U/r, \
{}^{(3)}\Gamma^{\phi}_{\phi \theta} = \cot \theta,
\end{split}
\end{align}
and the non-zero Levi-Civita connections of the metric (\ref{tortoise_at_univ}) are
\begin{align}
\begin{split}
&\Gamma^{\tau}_{\tau \tau} = - \frac{V}{2r^2 U},~
\Gamma^{\tau}_{\tau r^{\ast}} = \frac{1}{2r^2 U},~
\Gamma^{\tau}_{r^{\ast} \tau} = \frac{1}{2 r^2 U},~
\Gamma^{\tau}_{r^{\ast} r^{\ast}} = \frac{V'}{U^2},~
\Gamma^{\tau}_{\theta \theta} = \frac{r V}{U},~\\
&\Gamma^{\tau}_{\phi \phi} = \frac{r\sin^2 \theta V}{U},~
\Gamma^{r^{\ast}}_{\tau \tau} = \frac{1-1/r}{2 r^2 U},~
\Gamma^{r^{\ast}}_{\tau r^{\ast}} = \frac{V}{2 r^2 U},~
\Gamma^{r^{\ast}}_{r^{\ast} \tau} = \frac{V}{2 r^2 U},~
\Gamma^{r^{\ast}}_{r^{\ast} r^{\ast}} = \frac{V V'}{U^2},\\
&\Gamma^{r^{\ast}}_{\theta \theta} = \frac{1-r}{U},~
\Gamma^{r^{\ast}}_{\phi \phi} = \frac{(1-r) \sin^2 \theta}{U},~
\Gamma^{\theta}_{r^{\ast} \theta} = \frac{U}{r},~
\Gamma^{\theta}_{\theta r^{\ast}} = \frac{U}{r},~
\Gamma^{\theta}_{\phi \phi} = - \cos \theta \sin \theta,\\
&\Gamma^{\phi}_{r^{\ast} \phi} = \frac{U}{r},~
\Gamma^{\phi}_{\theta \phi} = \cot \theta,~
\Gamma^{\phi}_{\phi r^{\ast}} = \frac{U}{r},~
\Gamma^{\phi}_{\phi \theta} = \cot \theta,
\end{split}
\end{align}
where a prime denotes the derivative with respect to $r^{\ast}$. The quadratic and cubic laplacians can be computed directly from (\ref{lap})
\begin{align}
\Delta^2 &= \partial_{r^{\ast}}^4 + A(r) \partial_{r^{\ast}}^3 + B(r) \partial_{r^{\ast}}^2 +C(r) \partial_r + D(r),\\
\begin{split}
\Delta^3 & = \partial_{r^{\ast}}^6 + \left( A + \frac{2U}{r} \right)\partial_{r^{\ast}}^5 + \left( 2A' + B + \frac{2U}{r} A - \frac{\ell (\ell + 1)}{r^2} \right) \partial_{r^{\ast}}^4\\
&+ \left( 2 B' +C + \frac{2U}{r} A' + \frac{2U}{r} B - \frac{\ell (\ell+1)}{r^2} A + A'' \right) \partial_{r^{\ast}}^3\\
&+ \left( B'' +2C' + \frac{2U}{r} B' + \frac{2 U}{r} C - \frac{\ell (\ell + 1)}{r^2} B+D \right) \partial_{r^{\ast}}^2\\
&+ \left( C'' + \frac{2U}{r} C'-\frac{\ell (\ell+1)}{r^2} C +2D' + \frac{2UD}{r} \right) \partial_{r^{\ast}}\\
&+D'' + \frac{2U}{r} D' - \frac{\ell (\ell + 1)}{r^2} D,
\end{split}
\end{align}
where
\begin{align}
A &= \frac{4 U}{r}, \\
B &= \frac{4 U'}{r} - \frac{2 \ell (\ell + 1)}{r^2},\\
C&= - \frac{2}{r} \left( \frac{U U'}{r}- U'' \right),\\
D&=\frac{\ell (\ell + 1)}{r^2} \left( \frac{\ell (\ell + 1)}{r^2} - \frac{2 U^2}{r^2} + \frac{2 U'}{r} \right).\label{def_D}
\end{align}

\section{The convergence and consistency of numerical solutions}
\label{app_conv}
\begin{figure}[h]
\centering
    \includegraphics[width=0.6\textwidth]{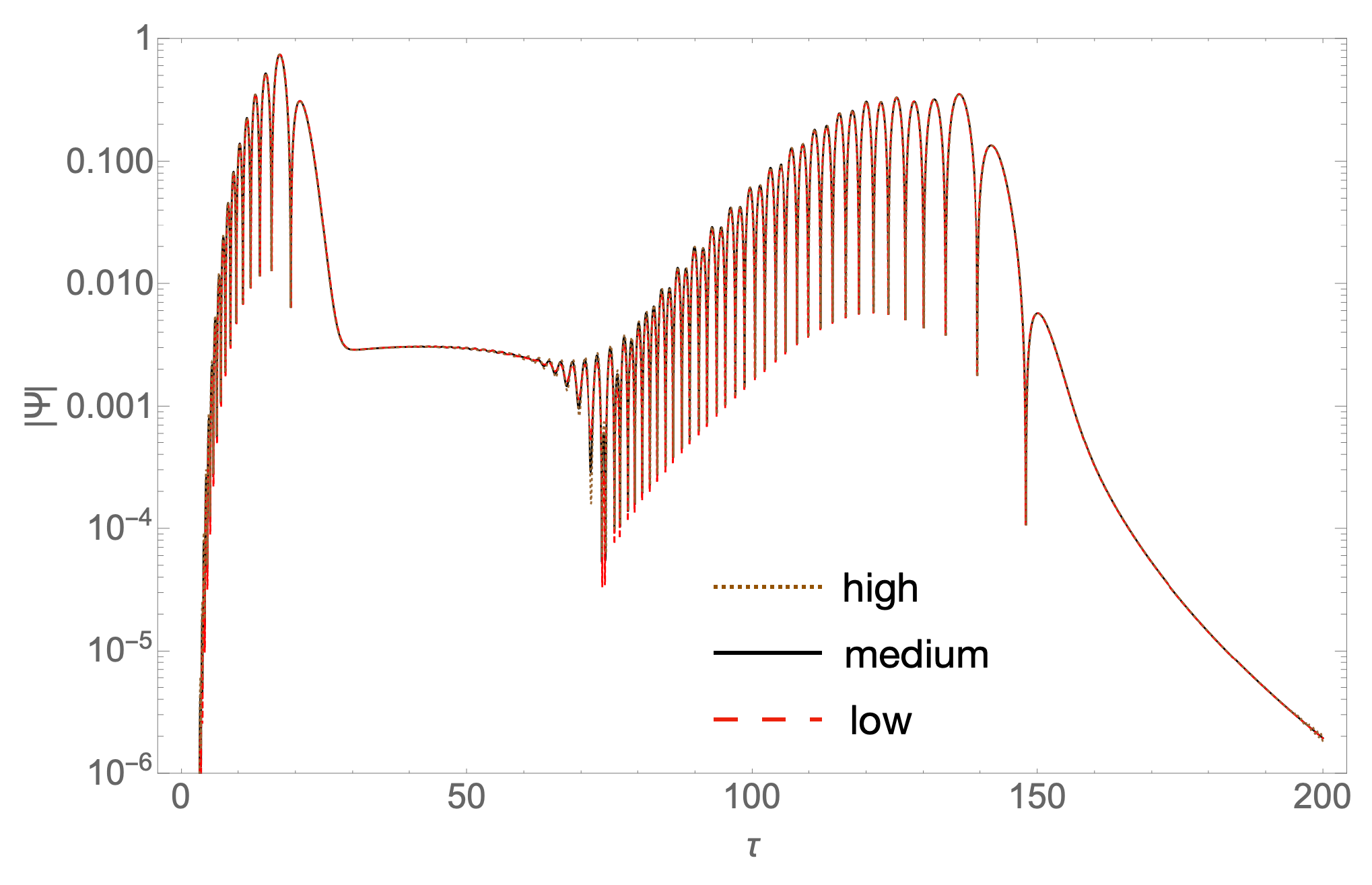}
\caption{Comparison among the results with $(\Delta \tau, \Delta r^{\ast}) =(0.008,0.16)$, $(0.021,0.21)$, and $(0.0385,0.256)$. The ratio $\lambda \equiv \Delta \tau/\Delta r^{\ast}$ is $0.05$, $0.1$, and $0.15$, respectively. The coefficient of the Kreiss-Oliger dissipation is $1/16$ and we use $\Xi_4=0.1$, $\Xi_6=0.01$, and $c_{\chi} = 0$.
}
\label{resolution}
\end{figure}

\begin{figure}[h]
\centering
    \includegraphics[width=0.6\textwidth]{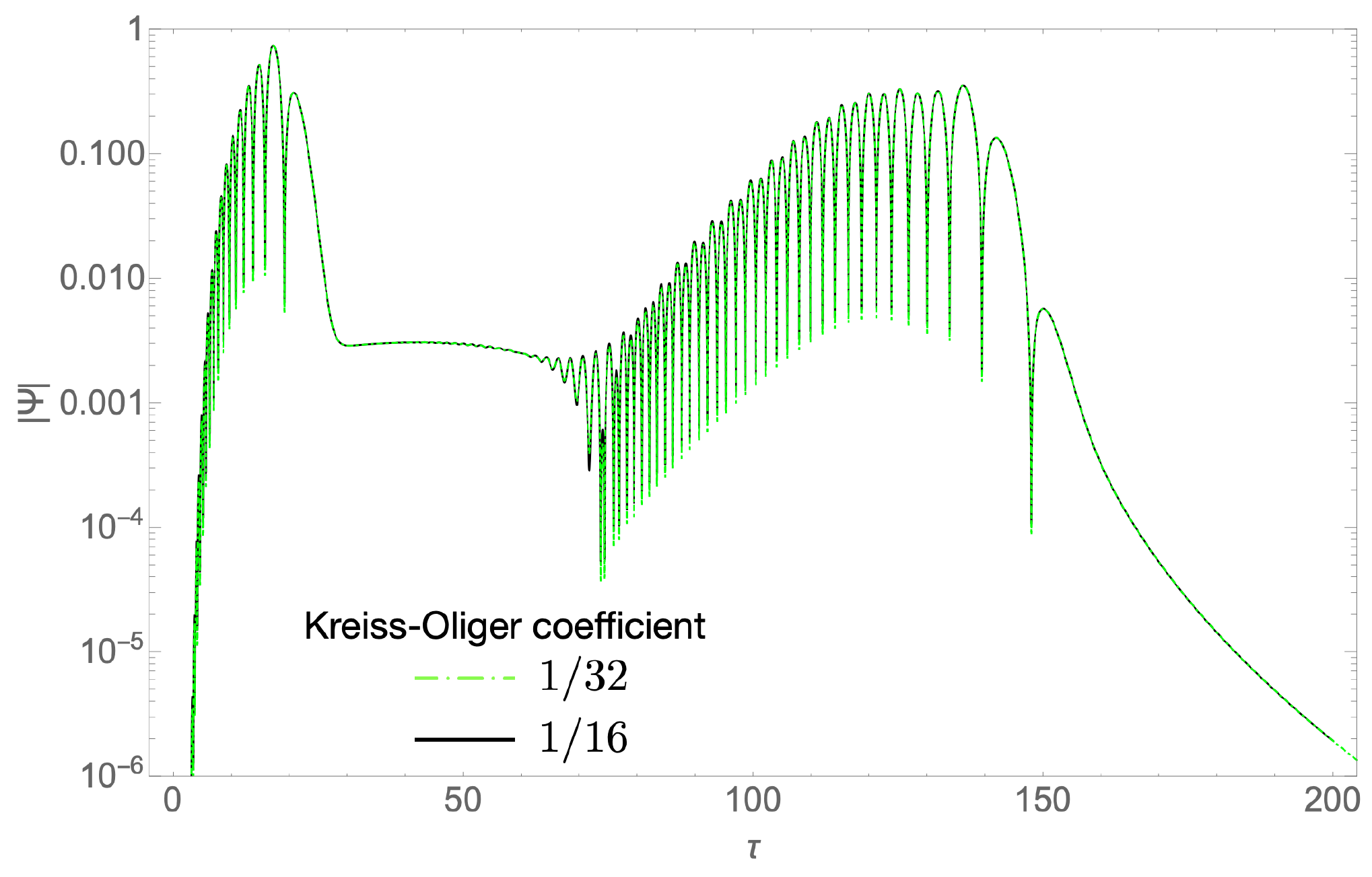}
\caption{The numerical simulation with the Kreiss-Oliger coefficient of $1/32$ and $1/16$. We use $\Xi_4=0.1$, $\Xi_6=0.01$, and $c_{\chi} = 0$.
}
\label{KO}
\end{figure}

\begin{figure}[h]
\centering
    \includegraphics[width=1\textwidth]{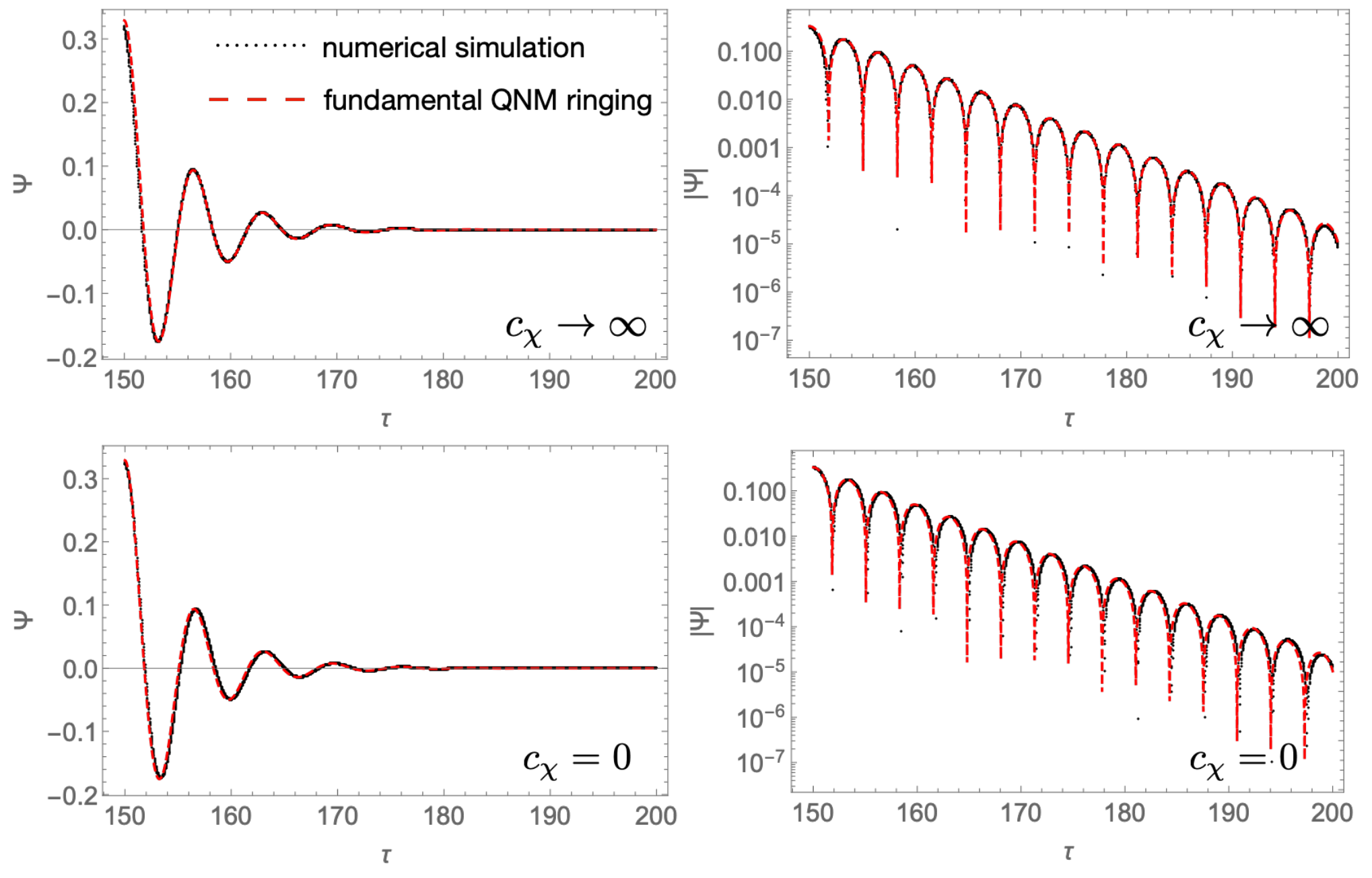}
\caption{The ringdown waveform of $\Xi_6 = \Xi_4 = 0$ computed by our numerical computation with $c_{\chi} \to \infty$ (top), and $c_{\chi} = 0$ (bottom). The red solid lines are the ringdown waveform obtained from the massless scalar fundamental quasinormal mode for $\ell = 2$.
}
\label{check}
\end{figure}

\begin{figure}[h]
\centering
    \includegraphics[width=0.7\textwidth]{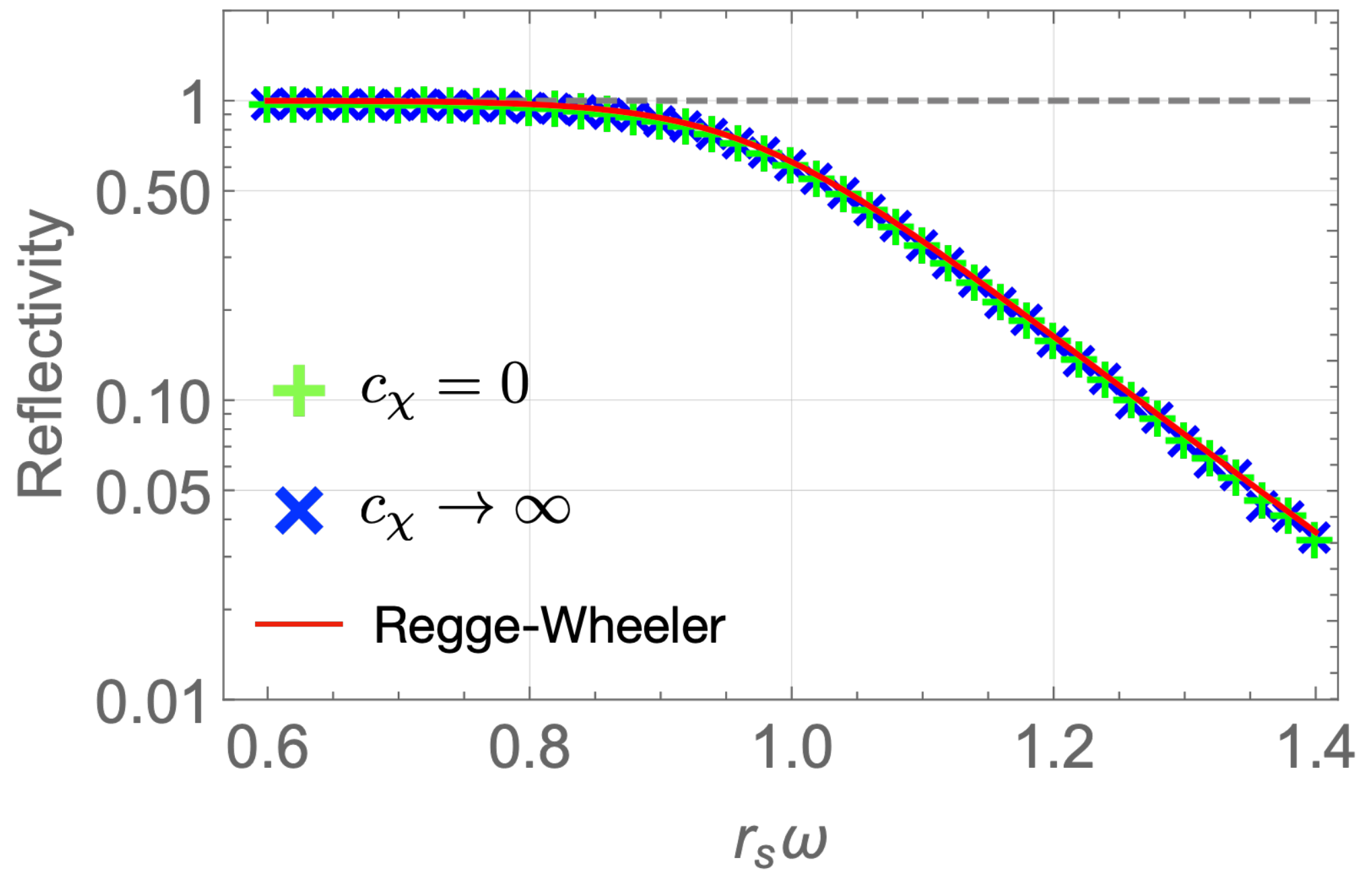}
\caption{The reflectivity of $\Xi_6 = \Xi_4 = 0$ computed by our numerical computation with $c_{\chi} \to \infty$ (cross) and $c_{\chi} = 0$ (plus). The red solid line is obtained from the numerical solution of the Regge-Wheeler equation.
}
\label{ref_check}
\end{figure}

The convergence of our simulations is tested by changing the resolution. We performed the numerical simulations with $(\Delta \tau, \Delta r^{\ast}, \lambda) =(0.008,0.16, 0.05)$, $(0.021,0.21,0.1)$, and $(0.0385,0.256,0.15)$ and one can find that the waveform converges well (FIG. \ref{resolution}). We also confirmed the Kreiss-Oliger dissipation does not affect the numerical result by performing our numerical simulation with different coefficients (FIG. \ref{KO}).
As a consistency check, we check that our numerical simulation reproduces the fundamental quasinormal mode at a late time when $M_{\rm HL} \to \infty$ (FIG. \ref{check}). The fundamental mode for a massless scalar field with $\ell = 2$ is $\omega_{\rm qnm} \simeq 0.9673 - i 0.1935$\footnote{The list of quasinormal modes is presented in \href{https://pages.jh.edu/~eberti2/ringdown/}{https://pages.jh.edu/~eberti2/ringdown/} and \href{https://centra.tecnico.ulisboa.pt/network/grit/files/ringdown/}{https://centra.tecnico.ulisboa.pt/network/grit/files/ringdown/}.}, and our result is well consistent with the fundamental mode. We performed the simulation with $\ell = 2$ mode. Also, the reflectivity we obtained from the simulation for $\Xi_4=\Xi_6=0$ is consistent with the solution of the Regge-Wheeler equation \cite{Regge:1957td} for a massless scalar field (FIG. \ref{ref_check}).

\section{Snap shots of the Lifshitz scalar waves}
\label{app_snap}
\begin{figure}[H]
\centering
    \includegraphics[width=1\textwidth]{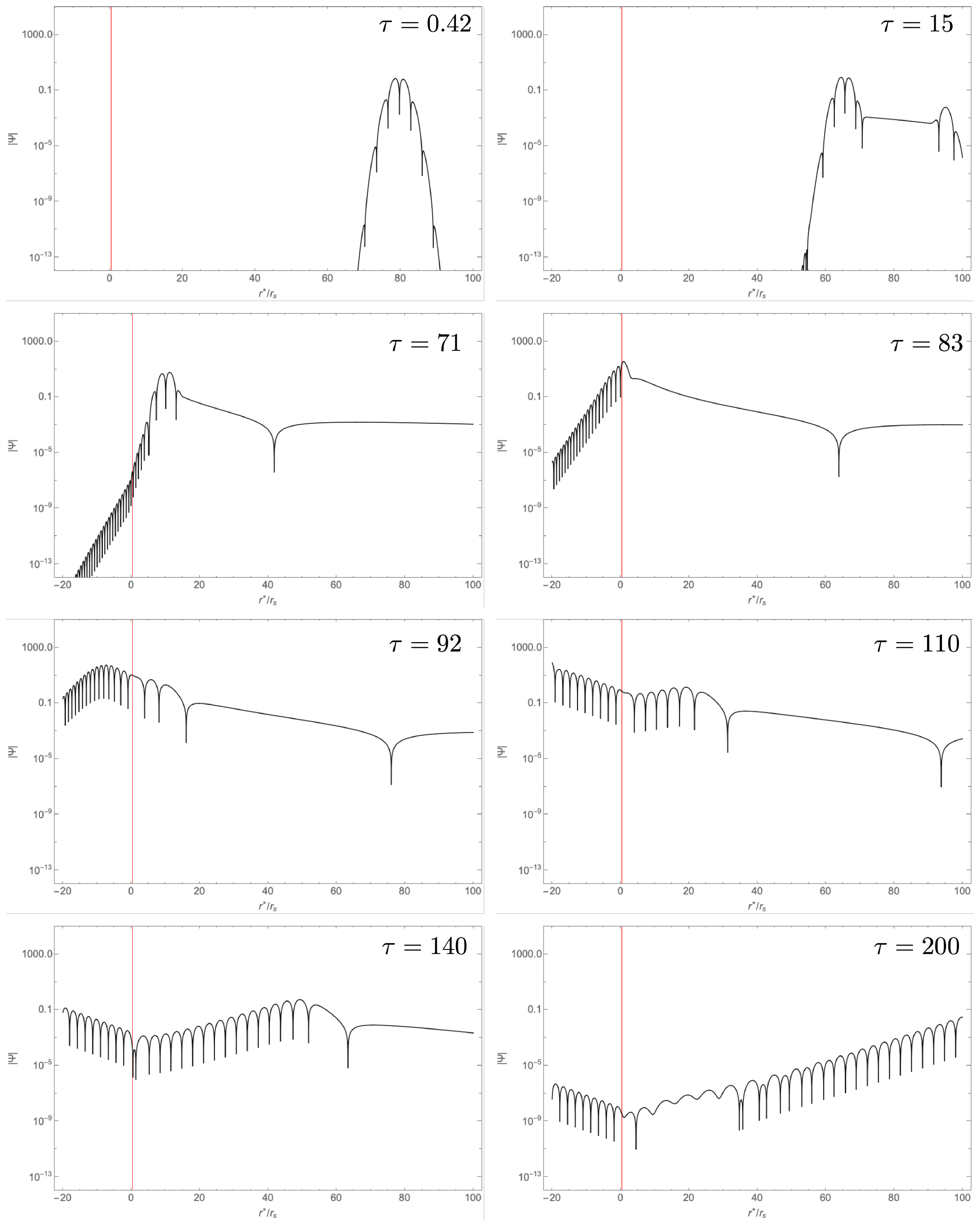}
\caption{Snap shots of the propagating scalar waves without the Lifshitz scaling. The red lines represent the position of the Killing horizon and we use $r^{\ast}_w = 80$ and $r^{\ast}_o = 60$. The universal horizon is located at $r^{\ast} \to -\infty$.
}
\label{snap_zero}
\end{figure}

\begin{figure}[H]
\centering
    \includegraphics[width=1\textwidth]{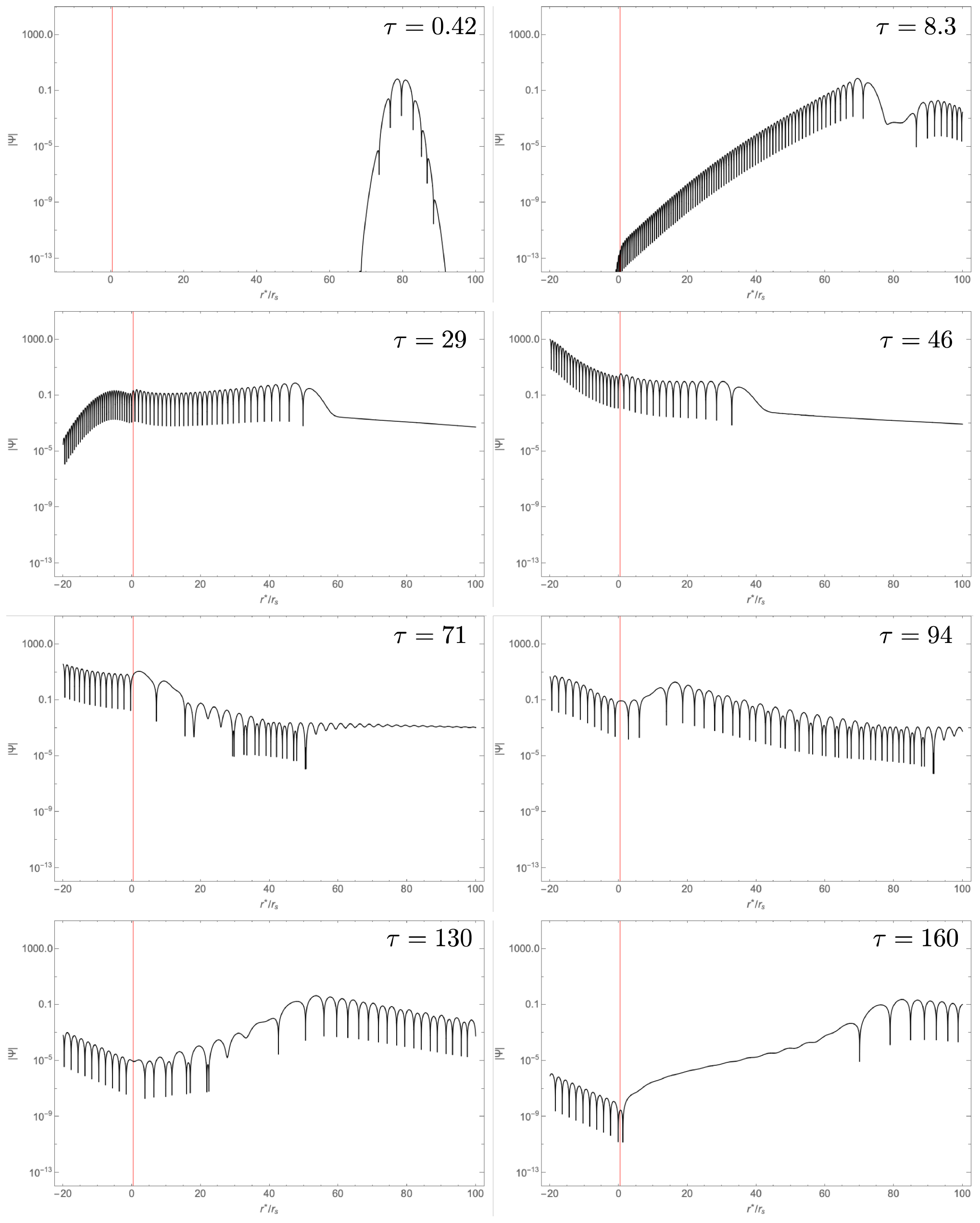}
\caption{Snap shots of the propagating Lifshitz scalar waves with $\Xi_4 = 0.1$, $\Xi_6 = 0.01$, and $c_{\chi} \to \infty$. We use $r^{\ast}_w = 80$ and $r^{\ast}_o = 60$.
}
\label{snap_posi}
\end{figure}

\begin{figure}[H]
\centering
    \includegraphics[width=1\textwidth]{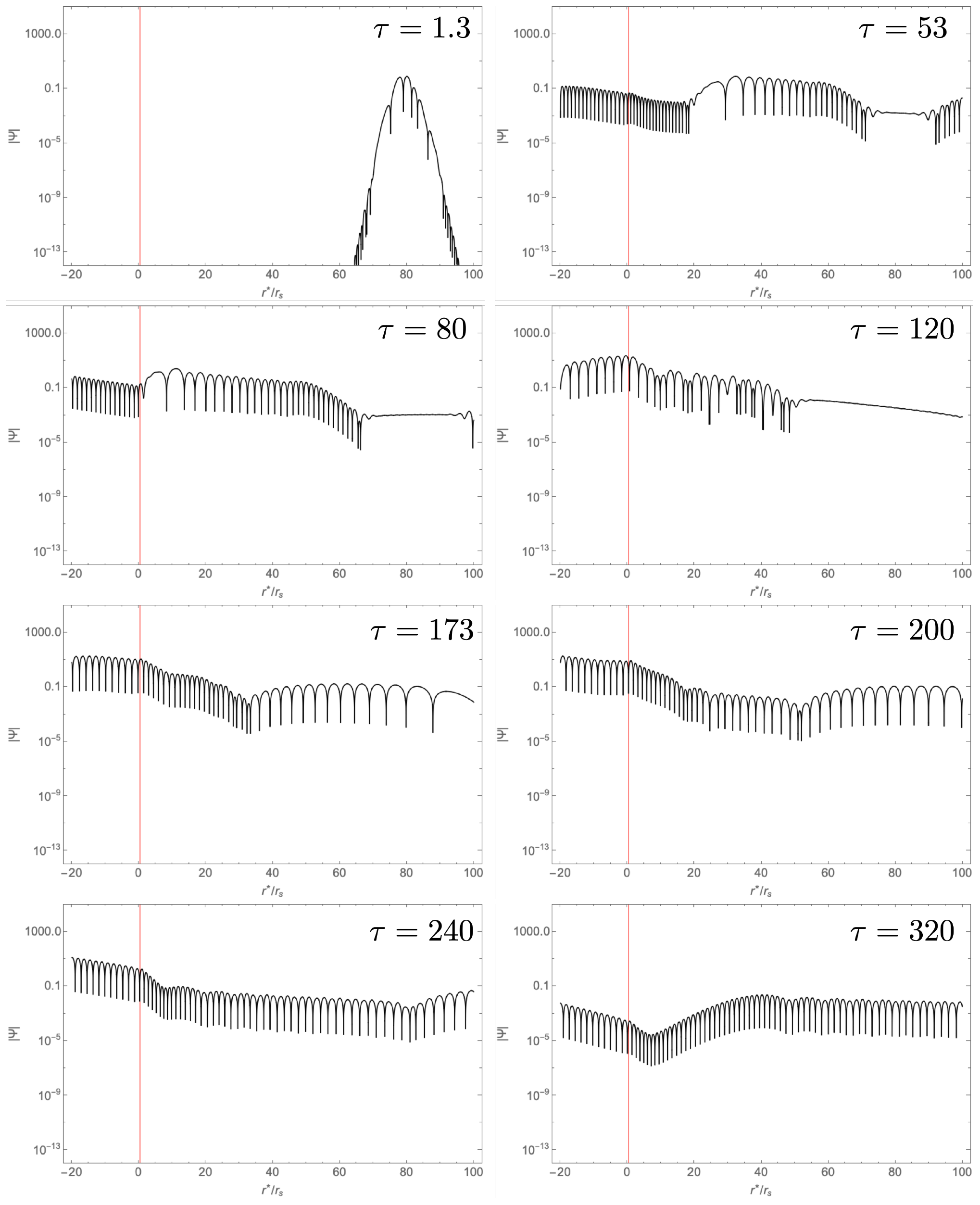}
\caption{Snap shots of the propagating Lifshitz scalar waves with $\Xi_4 = -0.15$, $\Xi_6 = 0.01$, and $c_{\chi} \to \infty$. We use $r^{\ast}_w = 80$ and $r^{\ast}_o = 60$.
}
\label{snap_nega}
\end{figure}

Here we show some snap shots of the perturbations of the Lifshitz scalar waves for three parameter sets: $(\Xi_4, \Xi_6) = (0,0)$ (FIG. \ref{snap_zero}), $(0.1,0.01)$ (FIG. \ref{snap_posi}), and $(-0.15,0.01)$ (FIG. \ref{snap_nega}). Although we present the snap shots only for $c_{\chi} \to \infty$, the trend does not change for the case of $c_{\chi} = 0$.
\newpage
\newpage
\bibliography{reference}
\end{document}